\documentstyle[psfig,conf_iap,10pt]{article}
\begin{document}
\heading{%
Far UV observation of a quasar sample
} 
\par\medskip\noindent
\author{%
Sandrine Fa\"{\i}sse$^{1}$, Bruno Milliard$^{1}$
}
\address{%
Laboratoire d'Astronomie Spatiale, BP 8,
13376 Marseille cedex 12, France
}
\begin{abstract}
The spectral indexes distribution of a quasar sample is derived from a far-UV 
survey. A red
average spectrum is found which comes in support of significant evolution of
the QSO spectral energy distribution (SED) since $z\approx2.5$
\end{abstract}
\\
\\
Photometry of quasars in the far-UV conveys information on their spectral
index at shorter wavelengths, and on the intergalactic medium (IGM)
opacity. Both factors are crucial for the ionisation of the IGM and for the
interpretation of quasar counts in UV.
\section{Data}
We present quasar observations at 200~nm from the balloon-borne imager FOCA
(limiting UV mag 18.5). A UV sample has been drawn from the 
cross-identification of the UV detections with quasars in \cite{Ver} and 
\cite{Bor} in a total of 10 square degrees mostly at high galactic latitudes. 
The UV sample is restricted to quasars confirmed with slit spectroscopy and
${\rm M_{B}}\leq-22$. One third of the 86 catalogued QSOs in
the surveyed area are detected. Their redshifts span the 0.2-1.8 range except
one at 2.4.
\section{Color distribution and the UV spectral indexes}
The $UV-B$ color distribution at ${\rm z}\leq1.1$ is determined using a maximum 
likelihood approach that makes use of  both measured fluxes and upper limits 
\cite{Av}. 
The colors are found to span a rather wide range, from $-1$ to $-3.5$, 
with a large blue fraction ($\simeq 25\%$ have ${\rm UV}-{\rm B}\leq-2$).
This UV$-$B color distribution, little affected by intergalactic  
opacity, is converted into spectral indexes $\alpha_{uv}$ in the rest-frame FUV
from $UV-B=-0.856(2-\alpha_{uv})-0.74$, with $f_{\nu}\propto\nu^{-\alpha}$. 
The best estimate of the spectral indexes distribution is presented in Fig 1.
\begin{figure}[ht]
\centerline{\vbox
{\psfig{figure=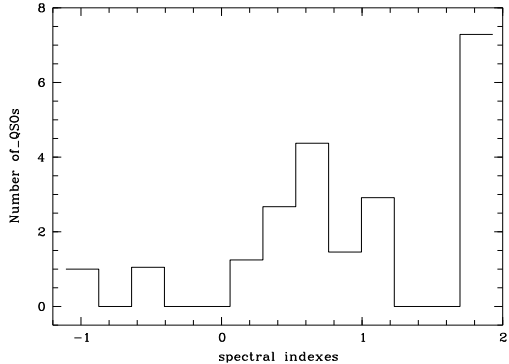,height=5cm,width=7.5cm,angle=-90}}}
\caption[]{Best estimate of the rest frame FUV spectral indexes
distribution.}
\end{figure}

The average value of the spectral indexes is $\alpha_{uv}=0.92$, in agreement
with that given by \cite{Zhe} from HST measurements. 
This comes in support of a softer rest frame FUV SED at lower z \cite{Fra}. 

The index distribution shows a significant dispersion (range [$-$1, 1.8],
$\sigma = 0.58$) compatible with that given
by \cite{Fra}, when some UV variability \cite{Dic} and
measurement errors are accounted for at the level of a few tenths of magnitude.
A higher red fraction than \cite{Fra} is found.
The bluest quarter of the obtained index
distribution corresponds to very blue continua, with an average index close
to 0.
\section{Predicted QSO counts in the UV}
We use the spectral index distribution derived above with current
estimates of the QSO luminosity function \cite{Laf}, \cite{War} and a 
simulation of the intergalactic opacity, to
anticipate the number of QSOs detectable by planned space UV experiments.
The effect of the intergalactic opacity is computed from a Monte-Carlo
simulation following \cite{Mol} but with current estimates of
the absorbers statistics and taking into account the damped Ly$\alpha$ 
systems. We assume H$_0$=50~km\,s$^{-1}$\,Mpc$^{-1}$~and~q$_0$=0.5.

The UV QSO counts are dominated by the bluest quarter of the index
distribution (indexes below 0.5). At m$_{2000}\le19$ and
${\rm M_B}\le-23$, we predict about 18~QSOs/sq deg, in agreement with the 6
objects found over 0.4 deg$^2$ in a genuine UV-selected sample in progress
at the Palomar 200-inches multifibre spectrograph. For an all-sky survey at
150nm limited to UV-magnitude 22, a number of 55~QSOs/sq~deg are expected.
%
%

%
\vfill
\end{document}